\documentclass{PoS}

\title{Full result for the three-loop static quark potential}

\ShortTitle{Three-loop static quark potential}

\author{Alexander V. Smirnov\\
Scientific Research Computing Center of Moscow State University, Russia\\
E-mail: \email{asmirnov80@gmail.com}}

\author{\speaker{Vladimir A. Smirnov}%
\\
Nuclear Physics Institute of Moscow State University, Russia\\
E-mail: \email{smirnov@theory.sinp.msu.ru}}

\author{Matthias Steinhauser\\
Institut f\"ur Theoretische Teilchenphysik, Karlsruhe
  Institute of Technology, Germany\\
E-mail: \email{matthias.steinhauser@kit.edu}}

\abstract{The three-loop corrections to the potential of two heavy quarks
  are computed. Analytic results for the most complicated master integrals are
  presented.}

\FullConference{RADCOR 2009 - 9th International Symposium on Radiative Corrections (Applications of Quantum Field Theory to Phenomenology) ,\\
		 October 25 - 30 2009\\
		 Ascona, Switzerland}

\newcommand{\ep}{\varepsilon}

\newcommand{\be}{\begin{equation}}
\newcommand{\ee}{\end{equation}}
\newcommand{\bea}{\begin{eqnarray}}
\newcommand{\eea}{\end{eqnarray}}

\newcommand{\gm}{\gamma}

\newcommand{\dd}{\mbox{d}}

\newcommand{\nn}{\nonumber}

\begin{document}

%- {{{ Introduction:

\section{Introduction}

The potential formed by two heavy quarks is among the early
applications of Quantum Chromodynamics (QCD). It is an important
ingredient in the description of the properties of
heavy-quark bound states. Among the most prominent applications, which
require high-order results of the static potential, are the top quark
production cross section close to threshold and the extraction of
the bottom quark mass from $\Upsilon$ sum rules (see, e.g.,
Ref.~\cite{Brambilla:2004wf} for a review).

The $n$-loop corrections
to the quark anti-quark potential
are usually parameterized by the constants $a_i$ such that
in momentum space it takes the form
\begin{eqnarray}
  V(|{\vec q}\,|)&=&
  -{4\pi C_F\alpha_s\over{\vec q}\,^2}
  \Bigg[1+{\alpha_s\over 4\pi}a_1
    +\left({\alpha_s\over 4\pi}\right)^2a_2
%    \nonumber\\&&\mbox{}
    +\left({\alpha_s\over 4\pi}\right)^3
    \left(a_3+ 8\pi^2 C_A^3\ln{\mu^2\over{\vec q}\,^2}\right)
    +\cdots\Bigg]\,,
  \label{eq::V}
\end{eqnarray}
where the renormalization scale has been identified
with $|{\vec q}\,|$ and for $SU(N_c)$ we have
$C_A=N_c$, $C_F=(N_c^2-1)/(2N_c)$ and $N_c=3$.

The one-loop corrections to $V(|{\vec q}\,|)$ have been computed more
than 30 years ago~\cite{1lp}. Around the
same time it has been discovered that  $V(|{\vec q}\,|)$ is not
infra-red safe and starting from three-loop
order divergences appears~\cite{Appelquist:1977es}.
In the 1990ies the two-loop corrections have been computed in
the works~\cite{Peter,Schroder:1998vy} and shortly
afterwards the coefficient of the three-loop divergence has been
evaluated
indirectly by examining the ultra-soft contribution to the energy of
two heavy quarks~\cite{Brambilla:1999qa,Kniehl:2002br}.
First steps towards the finite part of the three-loop corrections have
been performed in
Refs.~\cite{Smirnov:2008tz,Smirnov:2008pn} where the
fermionic contributions have been evaluated.
The gluonic contribution which completes
the knowledge about $a_3$ has been obtained by two
independent computations~\cite{Smirnov:2009fh,Anzai:2009tm}.

The evaluation of Refs.~\cite{Smirnov:2008pn,Smirnov:2009fh} is highly
automated in order to avoid errors due to manual interactions.
After generating the amplitudes for the Feynman diagrams with
{\tt QGRAF}~\cite{Nogueira:1991ex} we use
{\tt q2e} and {\tt exp}~\cite{Harlander:1997zb,Seidensticker:1999bb}
in order to rewrite the expressions in {\tt
  FORM}~\cite{Vermaseren:2000nd} format which is
used for taking the traces and further simplifications.
The reduction to master integrals is performed with
the program package {\tt FIRE}~\cite{FIRE} and the
resulting master integrals are computed with the help of the
Mellin-Barnes technique (see, e.g.,
Refs.~\cite{Smirnov:2004ym,Czakon,Smirnov:2009up}).
Cross checks are based on the program {\tt FIESTA}~\cite{FIESTA}
which incorporates the sector decomposition algorithm.

We managed to compute all the necessary
coefficients of the $\epsilon$ expansion of the master integrals
analytically with the exception of
three terms of order $\epsilon^1$.
Results for some master integrals are presented in the next
section. Let us close this Section by
summarizing the results for $a_3$ from Refs.~\cite{Smirnov:2008pn,Smirnov:2009fh}:
It is convenient to decompose $a_3$ according to the powers of $n_l$,
the numbers of massless quarks
%\begin{eqnarray}
  $a_3 = a_3^{(3)} n_l^3 + a_3^{(2)} n_l^2 + a_3^{(1)} n_l + a_3^{(0)}$
  \,.
%\end{eqnarray}
%The results of the individual coefficients, decomposed according to
%the colour structures, are given by:
These are our results:
\begin{eqnarray}
  a_3^{(3)} &=& - \left(\frac{20}{9}\right)^3 T_F^3
  \,,\nonumber\\
  a_3^{(2)} &=&
  \left(\frac{12541}{243}
    + \frac{368\zeta(3)}{3}
    + \frac{64\pi^4}{135}
  \right) C_A T_F^2
  +
  \left(\frac{14002}{81}
    - \frac{416\zeta(3)}{3}
  \right) C_F T_F^2
  \,,\nonumber\\
  a_3^{(1)} &=&
  \left(-709.717
  \right) C_A^2 T_F
%  \nonumber\\&&\mbox{}
  +
  \left(-\frac{71281}{162}
    + 264 \zeta(3)
    + 80 \zeta(5)
  \right) C_AC_F T_F
  \nonumber\\&&\mbox{}
  +
  \left(\frac{286}{9}
    + \frac{296\zeta(3)}{3}
    - 160\zeta(5)
  \right) C_F^2 T_F
%  \nonumber\\&&\mbox{}
  +
  \left(-56.83(1)
  \right) \frac{d_F^{abcd}d_F^{abcd}}{N_A} \,,
  \nonumber\\
  a_3^{(0)} &=&
  502.24(1) \,\, C_A^3
  -136.39(12)\,\, \frac{d_F^{abcd}d_A^{abcd}}{N_A}
  \,,
  \label{eq::a3}
\end{eqnarray}
where $T_F=1/2$,  $d_F^{abcd}d_F^{abcd}/N_A = (18 - 6 N_c^2 + N_c^4)/(96 N_c^2)$
and $d_F^{abcd}d_A^{abcd}/N_A = (N_c^3 + 6N_c)/48$.

%- }}}

\section{Results for selected master integrals}

\begin{figure}[b]
  \begin{center}
    \includegraphics[width=.7\textwidth]{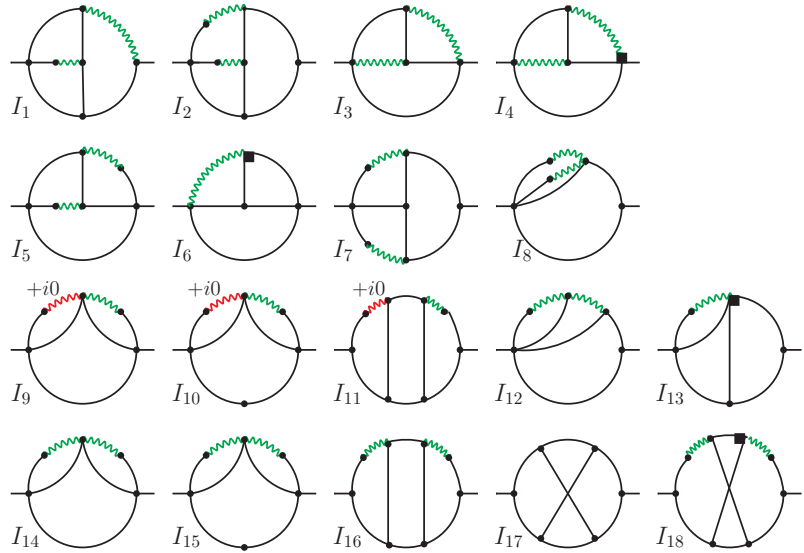}
   \caption[]{\label{fig::diags}Most complicated master integrals
%     needed for the evaluation
contributing to $a_3$. The solid and zig-zag lines
     correspond to relativistic and static propagators, respectively.
Small black boxes denote monomials in numerators.}
  \end{center}
\end{figure}

If one counts all master integrals for all the types
%\note{ms:  clear?, define?, rewrite?}
of the integrals appearing
in the calculation in the general $\xi$-gauge one obtains around hundred master
integrals. However, only 41 master integrals contribute to the three-loop static potential
in the Feynman gauge.
Eighteen most complicated master integrals
%where no explicit
%integration over some loop momentum in terms of $\Gamma$ functions can
%be performed
are shown in Fig.~\ref{fig::diags}.
Only two of them, $I_{17}$ and $I_{18}$, are non-planar.
Diagram $I_{17}$ does not involve static lines and is just a
three-loop propagator diagram which can be found in Ref.~\cite{npmincer}
and which we need up to order $\ep^1$.
Diagram $I_{18}$ represents the most complicated master integral in
our calculation.
Explicitly, we have $I_{18}=F^{(np)}_{1,\ldots,1,0,1,0}$, where
\bea
F^{(np)}_{a_1,\ldots,a_{12}}&=&\int\int\int\frac{\dd k\;\dd l\;\dd r}
{(-k^2)^{a_1}(-l^2)^{a_2}(-r^2)^{a_3}(-(r+q)^2)^{a_4}(-(k-l+r+q)^2)^{a_{5}}(-(k+q)^2)^{a_6}}
\nn \\  && \hspace*{-0mm} \times
\frac{(-(k-r)^2)^{-a_{12}}}
{(-(l-r)^2)^{a_7}(-(k-l)^2)^{a_8}(-v\cdot k)^{a_9}
(-v\cdot l)^{a_{10}}  (-v\cdot r)^{a_{11}}}\;,
\eea
with the causal $-i0$ implied in all propagators.
It turns out that it is more convenient to use,
instead of this master integral, a master integral with a numerator:
$I_{18}=F^{(\mbox{np})}_{1,\ldots,1,-2,1,0}$.
It is finite at $d=4$ and its value is one of three
(yet) analytically unknown constants.
The other two pieces of the three-loop static potential which are only
known numerically are the $O(\ep)$ terms in $I_{11}$ and $I_{16}$.

In the following we present our analytic results for the 16 most
complicated master integrals which are conveniently expressed as
special cases of one of the four functions
\bea
F^{(1)}_{a_1,\ldots,a_{12}}&=&\int\int\int\frac{\dd k\;\dd l\;\dd r}
{(-(k+q)^2)^{a_1}(-(r+q)^2)^{a_2}(-r^2)^{a_3}(-l^2)^{a_4}(-k^2)^{a_5}
(-(k-r)^2)^{a_6}}
\nn \\  && \hspace*{-0mm} \times
\frac{(-(l+q)^2)^{a_{12}}}
{(-(l-r)^2)^{a_7}(-(k-l)^2)^{a_8}(-v\cdot k)^{a_9}
(-v\cdot r)^{a_{10}}  (-v\cdot (k-l))^{a_{11}}}\;,
\\
F^{(2)}_{a_1,\ldots,a_{12}}&=&\int\int\int\frac{\dd k\;\dd l\;\dd r}
{(-(k+q)^2)^{a_1}(-(r+q)^2)^{a_2}(-r^2)^{a_3}(-l^2)^{a_4}(-k^2)^{a_5}}
\nn \\  && \hspace*{-0mm} \times
\frac{(-(l+q)^2)^{a_{12}}}
{(-(k-r)^2)^{a_6}(-(l-r)^2)^{a_7}(-(k-l)^2)^{a_8}(-v\cdot k)^{a_9}
(-v\cdot r)^{a_{10}}  (-v\cdot l)^{a_{11}}}\;,
\\
F^{(3)}_{a_1,\ldots,a_{12}}&=&\int\int\int\frac{\dd k\;\dd l\;\dd r}
{(-k^2)^{a_1}(-l^2)^{a_2}(-r^2)^{a_3}(-(r+q)^2)^{a_4}(-(l+q)^2)^{a_{5}}(-(k+q)^2)^{a_6}}
\nn \\  && \hspace*{-0mm} \times
\frac{(-(k-r)^2)^{-a_{12}}}
{(-(k-l)^2)^{a_7}(-(l-r)^2)^{a_8}(-v\cdot k)^{a_9}
(-v\cdot (k-l))^{a_{10}}  (-v\cdot r)^{a_{11}}}\;,
\\
F^{(4,\pm)}_{a_1,\ldots,a_{12}}&=&\int\int\int\frac{\dd k\;\dd l\;\dd r}
{(-k^2)^{a_1}(-l^2)^{a_2}(-r^2)^{a_3}(-(r+q)^2)^{a_4}(-(l+q)^2)^{a_{5}}(-(k+q)^2)^{a_6}}
\nn \\  && \hspace*{-0mm} \times
\frac{(-(k-r)^2)^{-a_{12}}}
{(-(k-l)^2)^{a_7}(-(l-r)^2)^{a_8}(-v\cdot k\mp i 0)^{a_9}
(-v\cdot l)^{a_{10}}  (-v\cdot r)^{a_{11}}}\;,
\eea
where in all propagators, apart \#9,
%from the propagator number 9 in $F^{(4,\pm)}$,
the causal $-i0$ is implied.
Here are our results:
%The analytical results for $I_1$ to $I_{16}$ read
\bea
%G(240,\{1,0,1,1,0,1,1,1,0,1,1,0\})
I_1=F^{(1)}_{1,0,1,1,0,1,1,1,0,1,1,0}&=&
\frac{1}{Q^2 v^2}
\left[
-\frac{28 \pi ^4}{135 \ep}
+\frac{188 \zeta (5)}{3}-\frac{152 \pi ^2 \zeta (3)}{9}
+\frac{112 \pi^4}{135} +O(\ep)
\right]\;,
\nn \\
%G(240,\{1,1,1,1,0,1,1,1,1,0,1,0\})
I_2=F^{(1)}_{1,1,1,1,0,1,1,1,1,0,1,0}&=&
\frac{1}{Q^4 v^2}
\left[
-\frac{44 \pi^4}{135 \ep}
-\frac{92 \zeta (5)}{3}-\frac{64 \pi ^2 \zeta (3)}{9}
-\frac{88 \pi ^4}{135} +O(\ep)
\right]\;,
\nn \\
%G(241,\{1,0,0,1,0,1,1,0,0,1,1,0\})
I_3=F^{(1)}_{1,0,0,1,0,1,1,0,0,1,1,0}&=&
\frac{\pi ^2 Q^2}{v^2}
\left[
-\frac{1}{9 \ep}
+\frac{4 \log (2)}{9}-\frac{41}{27}
\right.
\nn \\ &&  \hspace*{-30mm}
+\left.\left(-\frac{1201}{81}+\frac{43 \pi ^2}{108}+\frac{92 \log (2)}{27}
+\frac{8 \log^2(2)}{9}\right) \ep
\right.
%\nn \\  &&  \hspace*{-30mm}
+\left(-\frac{31289}{243}+\frac{1475 \pi ^2}{324} \right.
\nn \\  &&  \hspace*{-30mm}
\left.\left.+\frac{988 \log (2)}{81}
-\frac{11}{27} \pi^2 \log (2)+\frac{184 \log ^2(2)}{27}+\frac{32 \log ^3(2)}{27}
+\frac{85 \zeta (3)}{3}\right) \ep^2
+O(\ep^3)
\right]\;,
\nn \\
%G(241,\{1,-1,0,1,0,1,1,0,0,1,1,0\})
I_4=F^{(1)}_{1,-1,0,1,0,1,1,0,0,1,1,0}&=&
\frac{\pi ^2 Q^4}{v^2}
\left[
-\frac{7}{360\ep}
 +\frac{\log(2)}{10}-\frac{3041}{10800}
\right.
\nn \\ &&  \hspace*{-40mm}
 +\left(-\frac{909193}{324000}+\frac{37 \pi^2}{480}
   +\frac{2047 \log (2)}{2700}+\frac{\log ^2(2)}{5}\right) \ep
+\left(-\frac{238569389}{9720000}+\frac{114859 \pi ^2}{129600}
\right.
\nn \\  &&  \hspace*{-40mm}
\left.\left.
+\frac{78077 \log(2)}{27000}-\frac{11}{120} \pi ^2 \log (2)
+\frac{2047 \log ^2(2)}{1350}+\frac{4 \log^3(2)}{15}
+\frac{389 \zeta (3)}{72}\right) \ep^2
+O(\ep)^3
\right]\;,
\nn \\
%G(241,\{1,1,0,1,0,1,1,1,0,1,1,0\})
I_5=F^{(1)}_{1,1,0,1,0,1,1,1,0,1,1,0}&=&
\frac{1}{Q^2 v^2}
\left[
-\frac{32 \pi^4}{135 \ep}
-\frac{188 \zeta (5)}{3}-\frac{88 \pi ^2 \zeta (3)}{9}
+\frac{128 \pi ^4}{135}
+O(\ep)
\right]\;,
\nn \\
%G(250,\{0,1,0,1,0,1,1,1,1,-1,0,0\})
I_6=F^{(2)}_{0,1,0,1,0,1,1,1,1,-1,0,0}&=&
Q^2 \left[
\left(-\frac{5}{72}-\frac{\zeta(3)}{6}\right) \frac{1}{\ep}
-\frac{13 \zeta(3)}{9}-\frac{437}{432}
\right.
\nn \\   %\hspace*{-40mm}
&+&
\left.
\left(-\frac{23471}{2592}+\frac{5 \pi ^2}{288}-\frac{293 \zeta (3)}{27}
+\frac{\pi ^2 \zeta (3)}{24}+\frac{19 \zeta (5)}{6}\right) \ep
+O(\ep^2)
\right]\;,
\nn \\
%G(250,\{1,1,1,1,0,1,1,1,1,0,1,0\})
I_7=F^{(2)}_{1,1,1,1,0,1,1,1,1,0,1,0}&=&
\frac{1}{Q^4 v^2}
\left[
-\frac{32 \pi^4}{135 \ep}
+\frac{184 \zeta (5)}{3}-\frac{64 \pi ^2 \zeta (3)}{9}-\frac{64 \pi ^4}{135}
+O(\ep)
\right]\;,
\nn \\
%G(330,\{1,0,1,1,0,0,1,1,1,1,0,0\})
I_8=F^{(3)}_{1,0,1,1,0,0,1,1,1,1,0,0}&=&
\frac{1}{v^2}
\left[
\frac{2 \pi ^2}{9 \ep^2}
+\left(\frac{16 \pi^2}{9}+\frac{8 \zeta (3)}{3}\right)\frac{1}{\ep}
+\frac{112 \zeta (3)}{3}+\frac{61 \pi ^4}{270}+\frac{56 \pi ^2}{9}
\right.
\nn \\  && \hspace*{-40mm}
+\left(-\frac{416 \pi ^2}{9}+\frac{628 \pi^4}{135}+\frac{1184 \zeta (3)}{3}
+\frac{152 \pi ^2 \zeta (3)}{9}-\frac{232 \zeta (5)}{3}\right)\ep
%\nn \\  && \hspace*{-40mm}
+\left(-\frac{11104 \pi ^2}{9}+\frac{8534 \pi ^4}{135}
\right.
\nn \\  && \hspace*{-40mm}
\left.
+\frac{77647 \pi^6}{45360}+\frac{11200 \zeta (3)}{3}+\frac{2908 \pi ^2 \zeta (3)}{9}
%\right.
%\nn \\  && \hspace*{-40mm}
%\left.
\left.
-\frac{1240 \zeta(3)^2}{3}-\frac{2864 \zeta (5)}{3}\right) \ep^2
+O(\ep^3)
\right]\;,
\nn \\
%G(348,\{1,0,1,0,1,0,1,1,1,0,1,0\})
I_9=F^{(4,-)}_{1,0,1,0,1,0,1,1,1,0,1,0}&=&
\frac{1}{v^2}
\left[
-\frac{32 \pi ^4}{135 \ep}
   +\frac{226 \zeta (5)}{3}-\frac{232 \pi ^2 \zeta (3)}{9}
+4 \pi ^4 \log (2)-\frac{256 \pi^4}{135}
\right.
\nn \\  && \hspace*{-30mm}
+\left(-192 s_6+\frac{1808 \zeta(5)}{3}-\frac{8 \zeta (3)^2}{3}-\frac{1856 \pi ^2 \zeta (3)}{9}
-128 \pi ^2 \mbox{Li}_4\left(\frac{1}{2}\right)-\frac{16}{3} \pi ^2 \log ^4(2)
\right.
\nn \\  && \hspace*{-30mm}
+\left.\left.\frac{28}{3} \pi ^4 \log^2(2)+32 \pi ^4 \log (2)+\frac{2344 \pi ^6}{2835}
-\frac{2048 \pi ^4}{135}\right) \ep
+O(\ep^2)
\right]\;,
\nn \\
%(348,\{1,0,1,0,2,0,1,1,1,0,1,0\})
I_{10}=F^{(4,-)}_{1,0,1,0,2,0,1,1,1,0,1,0}&=&
\frac{1}{Q^2 v^2}
\left[
\frac{\pi ^4}{\ep}
-93 \zeta (5)+28 \pi ^2 \zeta (3)+2 \pi ^4 \log(2)
\right.
\nn \\  && \hspace*{-43mm}
+\left(-96 s_6+120 \zeta (3)^2-64 \pi^2 \mbox{Li}_4\left(\frac{1}{2}\right)
-\frac{8}{3} \pi ^2 \log ^4(2)
\right.
%\nn \\  && \hspace*{-30mm}
+\left.\left.\frac{14}{3} \pi ^4 \log^2(2)
+\frac{887 \pi ^6}{420}\right) \ep
+O(\ep^2)
\right]\;,
\nn \\
%(348,\{1,1,1,1,1,1,1,1,1,0,1,0\})
I_{11}=F^{(4,-)}_{1,1,1,1,1,1,1,1,1,0,1,0}&=&
\frac{1}{Q^6 v^2}
\left[
\frac{64 \pi ^4}{135 \ep}
-\frac{8 \zeta (5)}{3}+\frac{32 \pi ^2 \zeta(3)}{9}+\frac{128 \pi ^4}{135}
+O(\ep)
\right]\;,
\nn \\
%(350,\{1,0,1,1,0,0,1,1,1,1,0,0\})
I_{12}=F^{(4,+)}_{1,0,1,1,0,0,1,1,1,1,0,0}&=&
\frac{1}{Q^6 v^2}
\left[
\frac{2 \pi ^2}{9\ep^2}
-\frac{4 \zeta (3)}{3\ep} +\frac{16 \pi^2}{9\ep}
\right.
-\frac{56 \zeta (3)}{3}+\frac{157 \pi ^4}{270}+\frac{56 \pi ^2}{9}
\nn \\  && \hspace*{-40mm}
+\left(-\frac{416 \pi ^2}{9}+\frac{1156 \pi^4}{135}-\frac{592 \zeta (3)}{3}
+\frac{161 \pi ^2 \zeta (3)}{9}+\frac{116 \zeta (5)}{3}\right)\ep
%\nn \\  && \hspace*{-40mm}
+\left(-\frac{11104 \pi ^2}{9}+\frac{12758 \pi ^4}{135}
\right.
\nn \\  && \hspace*{-40mm}
\left.
+\frac{18563 \pi^6}{9072}-\frac{5600 \zeta (3)}{3}+\frac{3034 \pi ^2 \zeta (3)}{9}
%\right.
%\nn \\  && \hspace*{-40mm}
%\left.
+\left.\frac{620 \zeta(3)^2}{3}+\frac{1432 \zeta (5)}{3}\right) \ep^2
+O(\ep^3)
\right]\;,
\nn \\
%(350,\{1,0,1,1,1,0,1,1,1,-1,0,0\})
I_{13}=F^{(4,+)}_{1,0,1,1,1,0,1,1,1,-1,0,0}&=&
\frac{1}{6\ep^2}
+\frac{3}{2 \ep}
-18 \zeta (5)+18 \zeta(3)-\frac{\pi ^2}{24}+\frac{25}{6}
\nn \\  && \hspace*{-10mm}
+\left(-\frac{105}{2}-\frac{3 \pi ^2}{8}+\frac{3 \pi ^4}{5}-\frac{2 \pi^6}{63}
+\frac{619 \zeta (3)}{6}-50 \zeta (3)^2\right) \ep
+O(\ep^2),
\nn \\
%(350,\{1,0,1,0,1,0,1,1,1,0,1,0\})
I_{14}=F^{(4,+)}_{1,0,1,0,1,0,1,1,1,0,1,0}&=&
\frac{1}{v^2}
\left[
\frac{28 \pi ^4}{135\ep}
+\frac{116 \pi ^2 \zeta (3)}{9}
+\pi ^4 \left(\frac{224}{135}-4 \log (2)\right)
+\frac{226 \zeta (5)}{3}
\right. 
\nn \\  && \hspace*{-30mm}
+\left(-192 s_6+\frac{1808 \zeta (5)}{3}-\frac{8 \zeta(3)^2}{3}
+\frac{928 \pi ^2 \zeta (3)}{9}+64 \pi ^2
   \mbox{Li}_4\left(\frac{1}{2}\right)+\frac{8}{3} \pi ^2 \log ^4(2)
\right.
\nn \\  && \hspace*{-30mm}
   -\frac{20}{3} \pi ^4 \log^2(2)-32 \pi ^4 \log (2)
-\left.\left.\frac{428 \pi ^6}{2835}+\frac{1792 \pi ^4}{135}\right) \ep
+O(\ep^2)
\right]\;,
\nn \\
%(350,\{1,0,1,0,2,0,1,1,1,0,1,0\})
I_{15}=F^{(4,+)}_{1,0,1,0,2,0,1,1,1,0,1,0}&=&
\frac{1}{Q^2 v^2}
\left[
-\frac{\pi ^4}{\ep}
-93 \zeta (5)-14 \pi ^2 \zeta (3)-2 \pi ^4 \log(2)
\right.
\nn \\  && \hspace*{-40mm}
+\left.\left(-96 s_6+120 \zeta (3)^2+32 \pi ^2 \mbox{Li}_4\left(\frac{1}{2}\right)
+\frac{4}{3} \pi ^2 \log ^4(2)-\frac{10}{3} \pi ^4 \log^2(2)
-\frac{989 \pi ^6}{420}\right) \ep
+O(\ep^2)
\right]\;,
\nn \\
%(350,\{1,1,1,1,1,1,1,1,1,0,1,0\})
I_{16}=F^{(4,+)}_{1,1,1,1,1,1,1,1,1,0,1,0}&=&
\frac{1}{Q^6 v^2}
\left[
-\frac{56 \pi ^4}{135 \ep}
-\frac{8 \zeta (5)}{3}-\frac{16 \pi ^2 \zeta(3)}{9}-\frac{112 \pi ^4}{135}
+O(\ep)
\right]\;,
\nn
\eea
where $Q=\sqrt{-q^2}$ and
$(i\pi^{d/2}e^{-\gm_E \ep})^3$ is implied as a factor on the right-hand side.
These $\ep$-expansions are up to the order which contributes to the
static potential, with the exception of
$I_{11}$ and $I_{16}$, where one more order is desirable.
Results for all the master integrals as well as details of their calculation will be
published elsewhere.

{\em Acknowledgements.}
This work is supported by DFG through project SFB/TR~9
%``Computational Particle Physics''
and RFBR, grant 08-02-01451. V.S. appreciates the
financial support of the organizers of the Symposium.

\end{document}